# Parametric modelling of needle-tissue interaction using finite element analysis

KAUSTAV MONI BORA*, ADARSH MISHRA, AND CHERUVU SIVA KUMAR

Department of Mechanical Engineering, IIT Kharagpur, Kharagpur, West Bengal-721302, India

**Corresponding Author:**
Kaustav Moni Bora
School of Nano Science and Technology, IIT Kharagpur, West Bengal, India
Email: kstvbora@kgpian.iitkgp.ac.in, kstvbora@gmail.com

**Abstract** - The realistic modelling of medical interventions is the key requirement for development of high-fidelity medical simulators. In this work, a parameterized model suitable for real-time haptic feedback of needle interaction with tissue has been developed. The FE based numerical simulations are performed in ABAQUS/CAE and Comsol and then the results are post-processed in MATLAB to develop the parametric model. The obtained parametric model validated through comparison with various literature results. A MATLAB based GUI App has also been developed that utilizes these results and delivers an interface to examine the desired forces for any known needle position/orientation.

**Key Terms** – Tool tissue interaction, Parametric model, Tissue modelling, FE simulation, non-invasive interaction

## INTRODUCTION

One of the most important requirements for the development of high-fidelity simulators is realistic modelling of needle-tissue interactions. Simulators for surgery based on such models can provide a safe, efficient and ethical way for pre-operative planning, surgical training and practice. Surgical simulators based upon accurate modelling of human anatomy, and physiological response can be useful to medical students for complicated surgical practices. Since the patients in such simulators are virtual, patient safety concerns vanishes while students are learning. In addition to students, medical professional can also use such surgical simulators for pre- and intra-operative planning of surgical procedures.

Humans have constantly attempted to comprehend the anatomy's structure and function. It is critical to understand how surgical equipment interact with human tissues during medical procedures. Studies focused on such medical interventions are important for the design and development of medical tools that are meant to assist surgeons in specific surgical procedures. Around 800 B.C., in India, the earliest known surgical methods were reported by ayurvedic physician Sushruta, who described rhinoplasty (reconstruction of the nose with a flap of skin from forehead) and otoplasty (reconstruction of earlobe with skin from cheek) techniques in his medico-surgical compendium "Sushruta Samhita"[5]. Dissection was the only way to look inside a person's body before the invention of medical imaging. But with the advances in computer technology, it is now possible to even create simulations for surgical procedures. These simulators, when used with a haptic device, provide real-time viewing of the surgical operation as well as force feedback to the surgeon. Such a hysteroscopy training simulator is developed at ETH Zurich university in 2006[13].

Non-invasive procedures are surgical techniques that do not include tissue rupture. For modelling such non-invasive local and global needle-tissue interactions, several modelling strategies have been addressed in the literature. Changes in tool geometry generated variations in force-deflection responses only for large localised deformations of the material, according to Mahavash et al., 2002.[18]

Widely used method for modelling such interaction is finite element based linear elastic models[1,11,6,17]. Since isotropic and homogeneous materials required only two constants to define material behaviour;[2] models based on these linear elastic materials are easy to implement, have low computational cost and can be utilised for haptic rendering in real time. Since linear or non-linear elasticity-based FE models require huge computational capacity for real-time haptic simulation, many researchers have used optimized FE based techniques. One such optimized FE model for real-time surgical simulation has been developed by Bro-Nielsen et al.[3]. The optimization was based on the concept that the displacement of only those nodes that are near to tools are required. They have shown that results for nodal displacement from this condensed FE method are similar to that of conventional FE analysis. Cotin et al.[6] also created a modified FE method based real-time surgery simulator using bulk of computation during pre-processing stage of FE calculations.

The models given above are based on assumed material properties and have not been validated by experimental findings. However, utilising material properties determined from studies on actual or phantom tissues, several researchers have built linear elastic FE based models. DiMaio et al.[8] used non-invasive techniques to measure the tissue properties. Gosline et al.[12] developed a linear elasticity-based FE models and coupled to the haptic device used by DiMaio et al.[9] Kerdok et al.[16] established a method for determining the accuracy of a soft tissue model by comparing experimental research to FE models. A good agreement between experimental and FE simulations were observed in linear elasticity domain (stains less than 2%), however, at higher strains, these FE simulations did not match the experimental results well.

In surgical procedures, tissues undergo large deformations, to model these hyperelastic material models are required. A suitable strain energy function is necessary to create a hyperelastic model, from which the relationship between stress and strain can be obtainedIn the real-time training simulator, Wu et al.[21] used the Mooney-Rivlin model to simulate tissue deformation. To match the experimental data obtained from the small probe indentation results on perfused pig spleen, Davies et al.[7] used Neo-Hookean and Mooney-Rivlin hyperelastic models. Tie Hu et al.[15] conducted studies on pig liver and used Mooney-Rivlin and Ogden models to match the data. Chui et al.[5] also conducted studies on pig liver and used a combination of exponential, polynomial, and logarithmic strain energy functions to match the data.

A real soft tissue has both non-linear and visco-elastic properties. Puso and Weiss [20] have used an anisotropic visco-hyperelastic model, to simulate FE based model of the collateral ligament, using an exponential relaxation function to simulate quasi-linear visco-elastic behaviour.

To develop an effective surgical simulator, real-time information about the tissue deformation and corresponding needle reaction force is required. Generation of interactive simulation graphics corresponding to such interactions, while modelling physically correct behaviours of real human organs is quite a challenge. Production of video-like visualization of the needle-tissue interaction requires determination of this tissue deformation and corresponding reaction force at a rate greater than 20 times per second. As the human tissue is complex and behaves like a hyper elastic material, huge computation capacity would be required to develop a surgical simulator based on these parameters. For surgery simulation systems of non-critical organs, the speed and robustness of the models are considered to be more important than the accuracy. The need for real-time computation capacity in these simulators has inspired many researchers to develop or adapt simplistic models of elastic deformation for these surgical simulators. While these models have the advantage of being easy to implement and provide reasonably good computational speed, they lack precision and stability.

Doing real-time finite element analysis to simulate needle tissue-interactions requires very high processing capacity and it is not feasible to produce tissue deformation results at such high desired rate with this method. Hence, another technique has been developed to solve this problem. In this technique, a large number of FEA analyses have been performed for different positions and orientations of needle about a tissue sample. The results of these simulations are used to develop a parametric model that can be used to generate reaction forces corresponding to any given set of input related to position and orientation of needle with respect to tissue.

## MATERIALS AND METHODS

*Modelling and Simulation:*

*Tissue Modelling:*

A real tissue behaves as non-linear hyper-elastic material. However, due to complexity in modelling the hyper-elastic properties of the tissue and to reduce the computational cost associated with the hyper-elastic modelling in FE solvers, most of the researchers used linear or non-linear elastic material models of the tissue. In this work, however, a non-linear and hyper-elastic model of tissue is utilized. Out of many non-linear material models it has been observed that Ogden mode fits best for the rubber-like incompressible behaviour of real tissue.[13] Therefore, it is being used in the current research.

Ogden Model

The Ogden model's strain energy function is given by [15]:

$$U = \sum_{i=1}^{N} \frac{2\mu_i}{\alpha_i^2} (\lambda_1^{\alpha_i} + \lambda_2^{\alpha_i} + \lambda_3^{\alpha_i} - 3)$$

Where, U is the strain energy per unit of reference volume of the body (J mm$^{-3}$) and $\lambda_1$, $\lambda_2$ and $\lambda_3$ are the principal stresses (N mm$^{-2}$). Also, $\mu_i$ is the shear modulus (N mm$^{-2}$) in 'i' th direction and $\alpha_i$ is a dimensionless constant.

Thus, the stress ($\sigma$) vs. strain ($\varepsilon$) relationship for the Ogden model for Nth dimensional space is given by,

$$\sigma = \sum_{i=1}^{N} \frac{2\mu_i}{\alpha_i} (e^{-\varepsilon\alpha_i + \varepsilon} - e^{\frac{1}{2}(\varepsilon\alpha_i + \varepsilon)})$$

*Properties of the Tissue and the Needle*

The material properties for the tissue are taken from the Tie Hu et al. [14]. As most surgical instruments are made of stainless steel hence the needle in these simulations is assumed to be made of stainless steel.

Table 1: Material properties for needle and tissue

| Tissue | Needle |
|---|---|
| $\alpha_1$ = 11.085, $\mu_1$ = -0.288 (N mm$^{-2}$) | Young's modulus ($ET$): 210 GPa |
| $\alpha_2$ = 11.265, $\mu_2$ = 0.162 (N mm$^{-2}$) | Poisson's ratio: 0.3 |
| $\alpha_3$ = 10.996, $\mu_3$ = 0.067 (N mm$^{-2}$) | Material: Steel |

The aim of the simulation is to find the value of reaction force and reaction moments at the needle when the needle interacts with the tissue. It is also desirable to find out how the reaction force and the reaction moments vary with the position of needle, angle of orientation of needle and depth of penetration. In this work, a full 3D FE model is used to develop the parametric model.

Two basic parameters are considered while doing the simulation to find the reaction force and moments at the end of the needle.

- Position of the needle at the surface of the sample.
- Angle with the tissue, at which the needle interacting with the tissue sample.

Surgical needles generally have a wedge-shaped tip profile to provide easy penetration of intended tissue. The sharp edge of the needle initiate tissue rupture and the wedge profile helps in moving away the tissue material in its path. The needle used here, in the FE based simulations, has a sharp edge of 150 mm as shown in fig 1(b). The diameter and the length of the needle are taken as 1 mm and 60 mm respectively, to mimic a typical real needle used in the surgical procedures.

The tissue is selected to be a cuboid (fig 1(a)), the top surface of the tissue sizes 50x50 mm$^2$ and the depth is 30 mm. The dimensions of the tissue were selected on the basis of the experimental work performed by Tie Hu et al. [14], to be able to compare the results of the current simulations with those experimental results, and also considering the high computational cost associated with large number of elements in FE based simulation of a full 3D model.

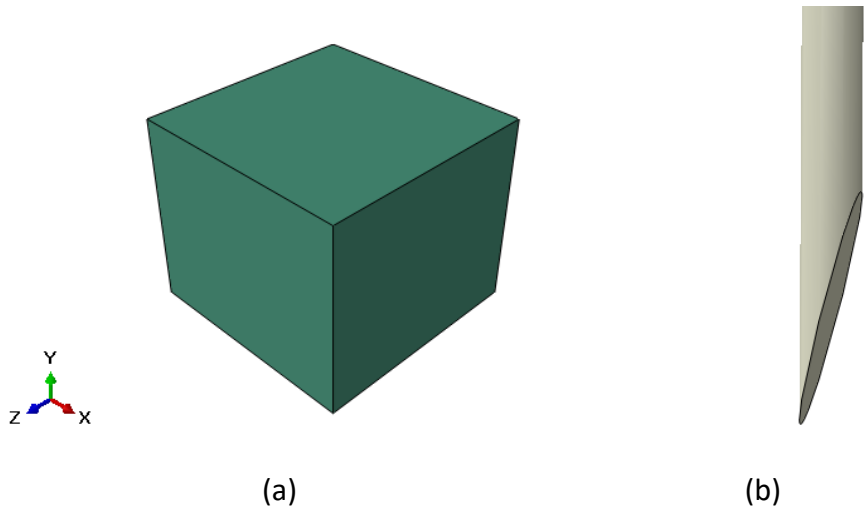

(a)          (b)

Figure 1: Geometrical model of tissue (a) and needle (b) for FE analysis

The tissue is meshed with 8-node linear hexagonal elements (C3D8) using structural meshing technique. The needle is meshed with same C3D8 elements and using a sweep meshing technique because of the complex geometry of the needle due to the presence of wedge-shaped tip.

*Boundary Conditions*

Following boundary conditions are applied on the tissue and needle on basis of the experimental work performed by Tie Hu et al. [14] and A. M. Okamura et al. [19]:

1. Sides of the tissue block are pinned to replicate the presence of a container.
2. Bottom surface of tissue block is encastred to replicate the presence of container surface.
3. Needle is feed a 6 mm of depth along the axis of needle in a total of 20 steps, each step of 0.3 mm.

*Constraints and Interactions*

Following interaction properties are assigned in the needle-tissue non-invasive model:

1. Initially, the top surface of the tissue lies in x-z plane, and the needle tip is placed on the centre of this surface, such that needle shaft is along the y-axis as shown in figure 3.
2. The contact between the needle and the tissue is modelled as 'Hard' in normal direction and 'frictional' in tangential direction, with the coefficient of friction equal to 0.5. [10]

3. A reference point is defined with respect to the needle shaft. This reference provides a mean to look for effective values of needle displacement and values of forces and moments acting on the needle.

*Development of Parametric Model:*

Multiple sets of FE based simulations are performed using same material properties of needle and tissue as mentioned in pervious section, but with different positions and orientation of needle as explained in the following sections.

*Variation of needle-tip in x-z plane*

As the top surface of the tissue sample is symmetrical (square), the variation of reaction force and reaction moments for different positions of the needle in both the axis will be same. Thus, simulations are done for different positions of the needle for one axis and that results are used for both the axes.

7 points are selected equidistant from each other from the centre of the tissue sample towards the centre of one of the edges and the simulations are performed for each such position. The distance between two consecutive points is 2 mm. The values of x (or z) coordinate at which needle is pointed are 0, 2, 4, 6, 8, 10 and 12 mm as shown in the figure 2(a).

*Variation of angle of inclination of needle*

Different set of simulations are made for variation of needle orientation in both x- and z- axis. The angles are taken as 0, 2, 4, 8, 12 and 15 degree with the y – axis in both positive and negative direction about x- and z- axes. Figure 2(b) shows one such typical orientation of the needle with $\theta$ angle in positive direction with respect to y axis.

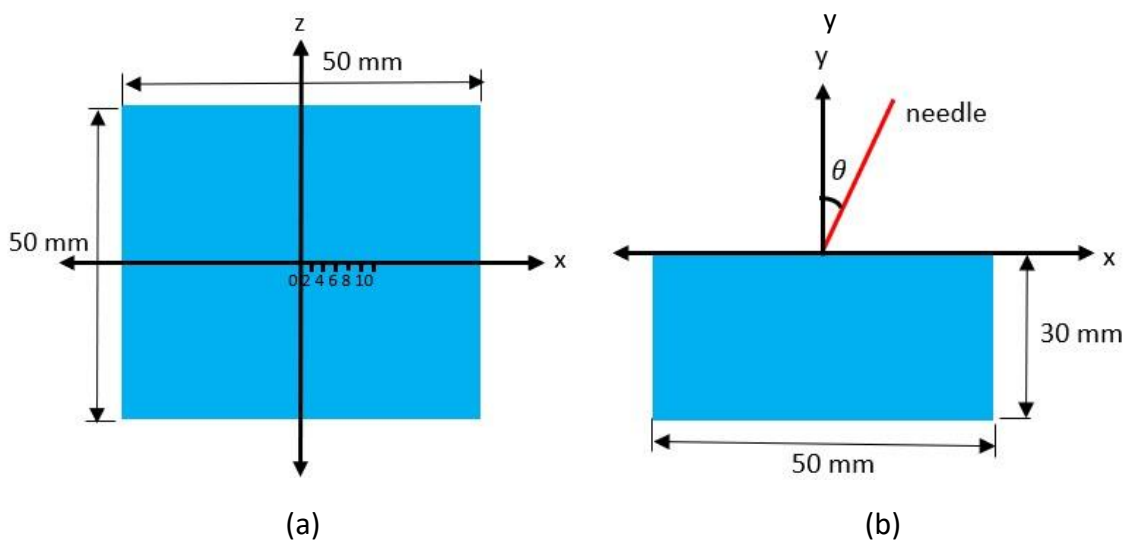

(a)          (b)

Figure 2: (a) Data points for variation of position of needle along x-axis (b) Orientation of the needle with $\theta$ angle in positive x- direction with respect to y axis

*Post-Processing in MATLAB:*

Getting data for individual reaction force component: The results from FE based simulations performed in ABAQUS/CAE (version 6.14) for different values of position and orientation on needle are imported in MATLAB. This complete input data is in form of 24 matrices as discussed in the result section. These data points are then re-arranged according to particular reaction force component i.e., each element of a matrix represents one particular reaction component. A total of 24 such matrices are generated, 4 for each reaction force components ($F_X$, $F_Y$, $F_Z$, $M_X$, $M_Y$ and $M_Z$).

*Finding parameters for variation of reaction force components*

For each reaction force component first its variation with depth of penetration was fitted using a 4th order polynomial, then again, the variation in these coefficients with position/orientation of needle is fitted using a 4th order polynomial. This operation was performed for all the 24 reaction force matrices. The results from these operations are in the form of 24 new 5x5 reduced coefficient matrixes.

*Calculation of effective reaction force at any given position and orientation*

To obtain the values of reaction forces for any given position/orientation of needle, first the coefficient matrix corresponding to variation in reaction forces with position in x- and z-direction are calculated from given values of x- and z- coordinate of needle position. Then, percentage change in the value of reaction forces due to angle of interaction of needle along x- or z-axis is calculated using corresponding coefficient matrices. The net reaction forces are then computed by superimposing these two results.

## RESULTS

The needle is placed at the centre of the top surface of the tissue and given a depth of penetration of 6 mm. Finite element-based simulations are performed for full 3D model of the needle-tissue interactions in ABAQUS/CAE (version 6.14) using the static implicit solver. The resulting deformation of the tissue and the corresponding reaction forces acting on the needle are shown in figure 3 and figure 4 respectively.

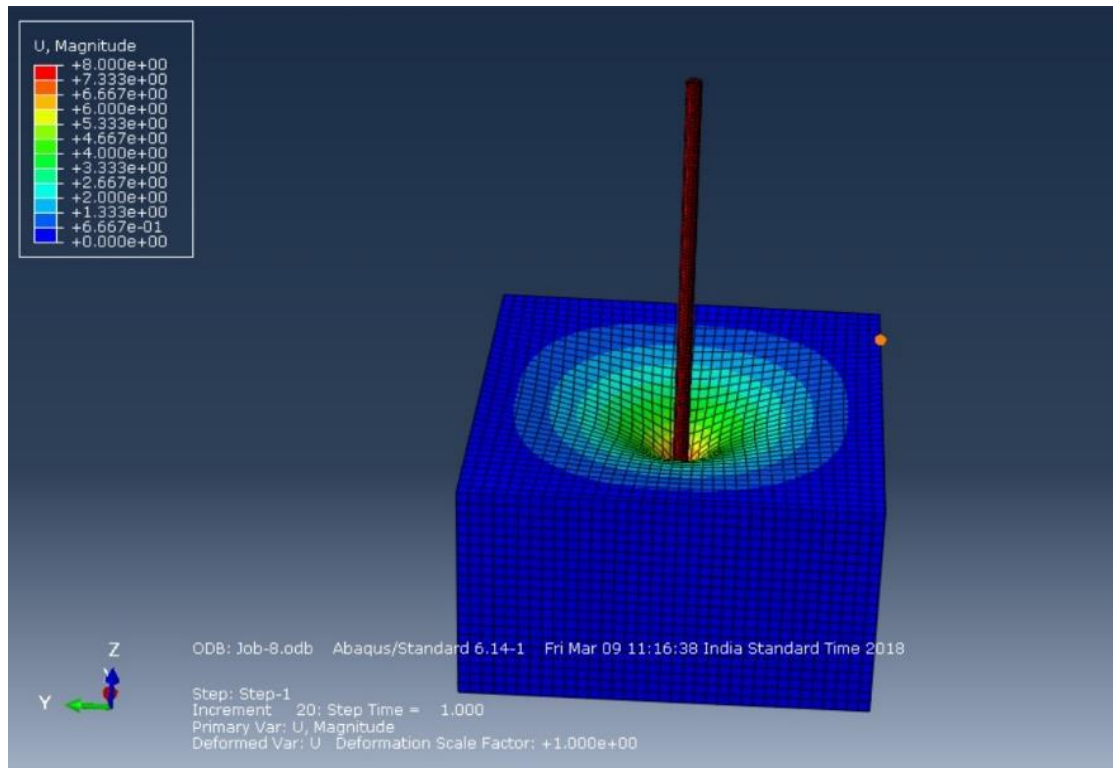

Figure 3: Tissue deformation in mm resulting from non-invasive needle interaction.

The reaction forces components ($F_X$, $F_Y$, $F_Z$, $M_X$, $M_Y$ and $M_Z$) resulting from non-invasive needle penetration up to 6 mm depth, are shown below in figure 4.

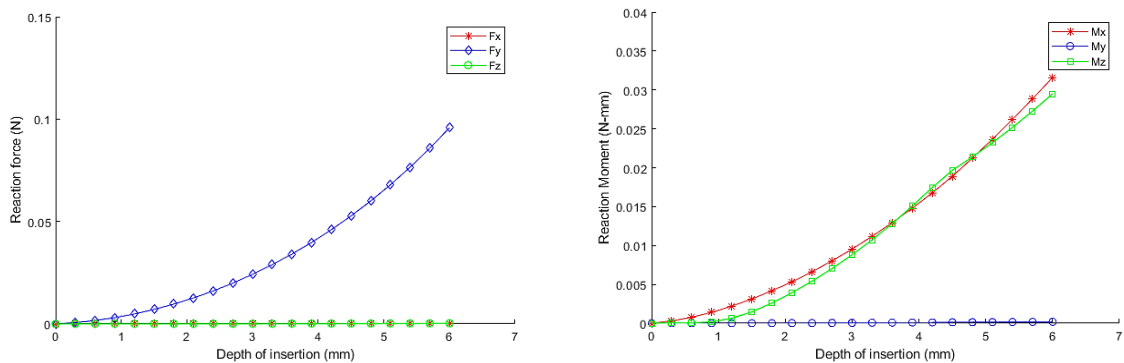

Figure 4: (a) Variation of reaction forces $F_X$, $F_Y$ and $F_Z$ with respect to depth of penetration of needle (b) Variation of reaction moments $M_X$, $M_Y$ and $M_Z$ with respect to depth of penetration of needle.

From the variation of reaction forces and moments shown in the figure, it can be seen that $Fy$ is the most predominant reaction forces component.

*Verification of FE simulated model:*

Results from the simulations are compared with reaction force values reported in experimental observations of Tie Hu et al.[14] and A. M. Okamura et al.[19] for purpose of verification.

The behaviour of the results is similar in both the simulated results and the reported experimental measurements collected by A. M. Okamura et al.[19]. The difference in the magnitude of the reaction forces is because of the fact that different tissues, and hence different material properties, were used in different research works.

*Results from Parametric Model*

The tissue deformation date at various location and with different angle of inclination are recorded. These data are exported from ABAQUS in text files for each unique position and orientation of the needle. Since in the simulation there are a total of 7 unique positions of needle in the x-z plane, and there are 6 unique angular orientations of needle at the top surface of the tissue, hence a total of 13 results are obtained. These results are imported in MATLAB for post-processing.

Each output matrix from the ABAQUS simulation contains:

    21 rows      – for different depth values (0 to 6 mm in 20 steps), and

    6 columns     – for 6 reaction components (3 forces and 3 moments).

One such typical tissue deformation for the inclined needle is shown in figure 6. Also, the results for variation in reaction force ($F_X$, $F_Y$ and $F_Z$) with respect to the variation in the position of the needle along x- and z-axes, and the variation in reaction force ($F_X$, $F_Y$ and $F_Z$) with respect to the variation in the orientation of the needle about x- and z-axes is shown in figure 6. Similarly, figure 7 shows the variation in reaction moments about for such variation in position and orientation of the needle.

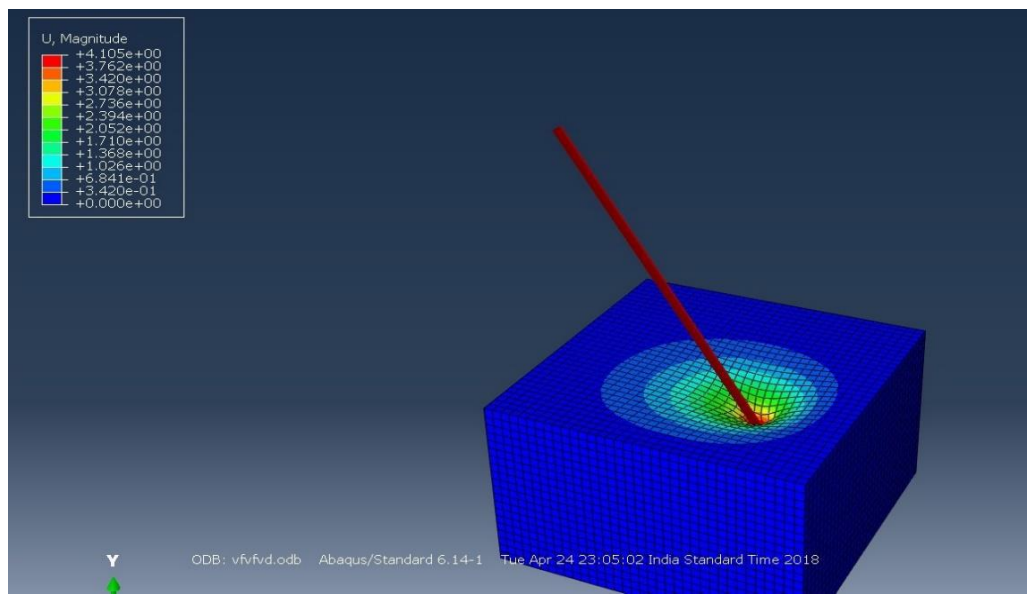

Figure 5: A typical simulation result for inclined needle at $x$ = 4 mm, z=0 and $\theta$=15°

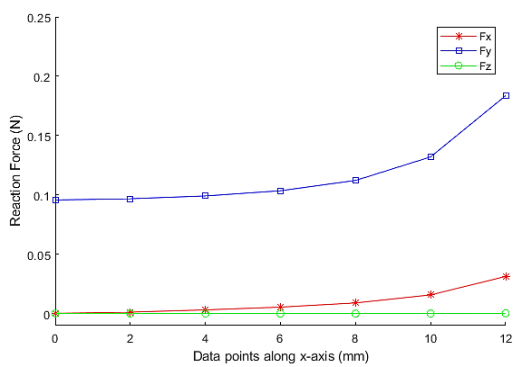 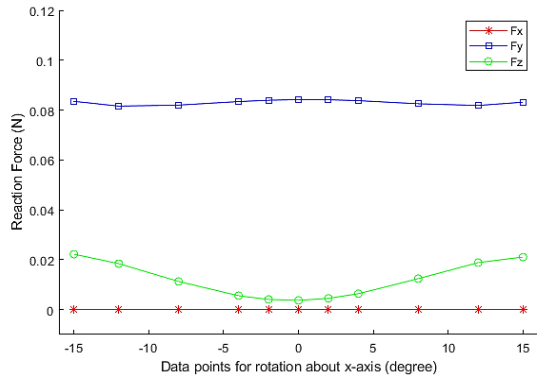

Figure 6: (a) Variation of reaction forces $F_X$, $F_Y$ and $F_Z$ with respect to variation in position of needle along x-axis, (b) Variation in reaction forces with respect to variation in orientation of needle about x-axis.

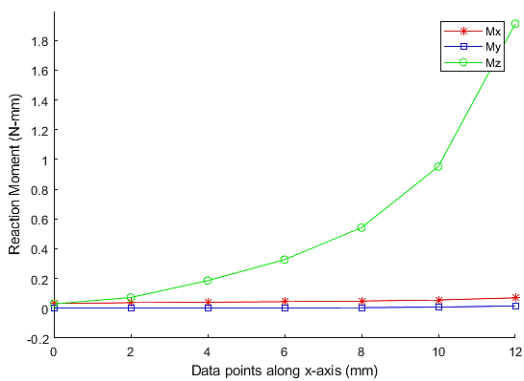 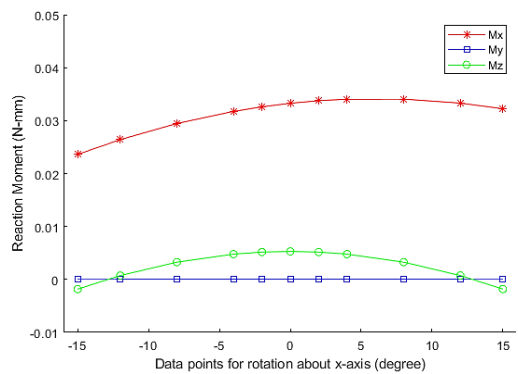

Figure 7: (a) Variation in reaction moments with respect to variation in position of needle along x-axis (b) Variation in reaction moments with respect to variation in orientation of needle about x- axis

*Validation of Parametric Model*

To check the validity of results obtained from the parametric model developed using MATLAB post-processing, some random position and orientation of the needle are assumed as in table 2. The value of reaction forces and moments are calculated using both, the parametric model, and by running FE simulation for these chosen positions and orientations of the needle.

Table 2: Position/orientation of the needle for the random points selected for the purpose of parametric model validation

| Position/orientation of the needle | Point_1 | Point_2 | Point_3 | Point_4 |
|---|---|---|---|---|
| Rotation about x-axis (in degree) | 0 | 4 | -4 | 7 |
| Rotation about z-axis (in degree) | 0 | 7 | 5 | -5 |
| Translation along x-axis (in mm) | -3 | 2 | 4 | -3 |
| Translation along z-axis (in mm) | 1 | 4 | 7 | 1 |

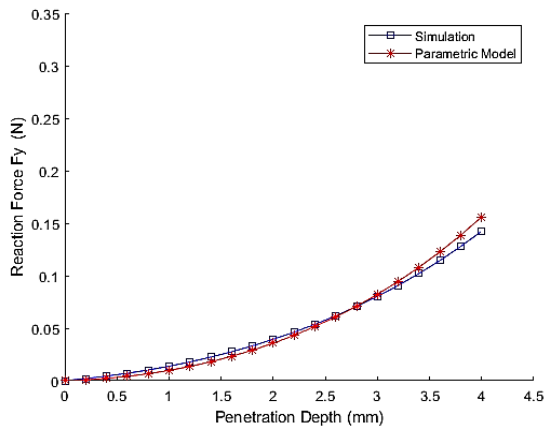

(a)

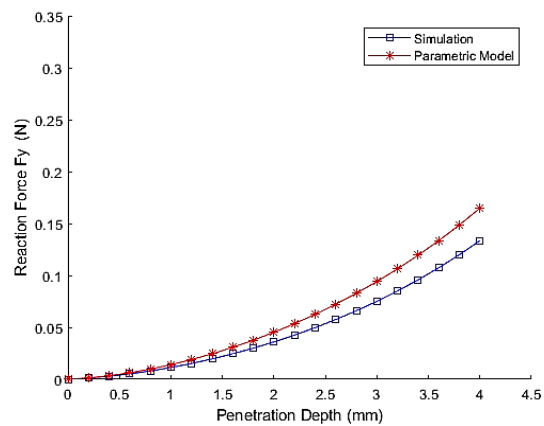

(b)

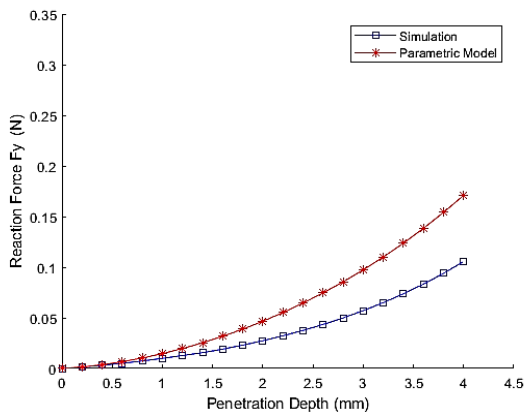

(c)

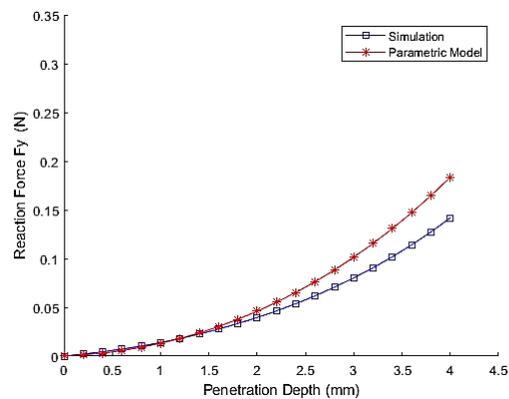

(d)

Figure 8: Comparison of the FE based simulations and parametric model results for reaction force component $F_Y$ for (Point 1 to Point 4) from tables

As mentioned earlier from figure 4, $F_Y$ is prominent force component, therefore it has been used here for the purpose of validation of parametric model, as shown in figure 8. It is observed that the results from the parametric model match closely with the simulation results.

The time required to do the simulation in ABAQUS and COMSOL is recorded and it is compared with the time required to compute the results from the parametric model. Both the simulations are performed in a same system (core i5 8$^{th}$ gen processor, 8 GB RAM) to make the comparison. The following results are obtained:

Time required to compute the simulation in ABAQUS or COMSOL = 81 s

Time required to compute the parametric model in MATLAB = 5.12 s

This enables the use of the results from the parametric model directly in a haptic system with fast rendering of solutions.

## DISCUSSION

The needle-tissue interaction forces are analysed in non-invasive phenomena. The results obtained from the FE simulations are verified by comparing with the reported experimental results. In the parametric model, it is seen that with increasing depth, the difference of values from the FE simulation and parametric model goes increasing. Also, as the operating point approaches boundary, both the model shows a large variation of reaction force. Thus, area of operation should be chosen considering this factor.

Data obtained from the palpation models are used to develop a parametric model that can be implemented in a real-time surgical simulator to provide information about the tissue deformation and the reaction forces at a very high rate.

In this work, the tissue is assumed to be hyper-elastic material and the material property of the tissue are taken from different literature already published. In future work inclusion experimental data of tissue could be taken into account. As the tissue properties varied from person to person, inclusion of non-linear visco-elastic and hyper-elastic nature of the tissue would provide more realistic results. In this work, the parametric model is developed using MATLAB. In future, this model can be implemented in a haptic-feedback device-based CHAI 3D to generate real-time haptic force interactions.

The MATLAB script used for these calculations can be available upon contacting authors.

## CONFLICT OF INTEREST

No benefits in any form have been or will be received from a commercial party related directly or indirectly to the subject of this manuscript.